\documentstyle[12pt]{article}

\begin{document}

\begin{titlepage}

\vspace*{-2cm}


\vspace{.5cm}

\begin{centering}

{\huge Brackets in the jet-bundle approach to field theory}

\vspace{1cm}

{\large G. Barnich}

\vspace{1cm}

Freie Universit\"at Berlin, Fachbereich Physik, Institut f\"ur
Theoretische Physik, Arnimallee 14, D-14195 Berlin.
\vspace{.5cm}

\vspace{1cm}

\begin{abstract}
In the first part
the sh Lie structure of brackets in field theory, described in the jet
bundle context along the lines suggested by Gel'fand, Dickey and
Dorfman, is analyzed. 
In the second part, we discuss how this description allows us to
find a natural relation between the Batalin-Vilkovisky antibracket and
the Poisson bracket. 
\end{abstract}

\end{centering}

\vspace{3cm}
{\footnotesize \hspace{-0.6cm}($^*$)Alexander-von-Humboldt fellow. 
On leave of absence from Charg\'e de Recherches du Fonds National
Belge de la Recherche Scientifique at Univerit\'e Libre de Bruxelles.}

\end{titlepage}

\pagebreak

\newtheorem{theorem}{Theorem}
\newtheorem{lemma}{Lemma}
\newtheorem{definition}{Definition}
\newtheorem{corollary}{Corollary}

\renewcommand{\theequation}{\thesection.\arabic{equation}}
\renewcommand{\thetheorem}{\thesection.\arabic{theorem}}
\renewcommand{\thelemma}{\thesection.\arabic{lemma}}
\renewcommand{\thecorollary}{\thesection.\arabic{corollary}}
\renewcommand{\thedefinition}{\thesection.\arabic{definition}}

\def\CJ{Loc(E)}
\def\be{\begin{eqnarray}}
\def\ee{\end{eqnarray}}
\def\qed{\hbox{${\vcenter{\vbox{                         
   \hrule height 0.4pt\hbox{\vrule width 0.4pt height 6pt
   \kern5pt\vrule width 0.4pt}\hrule height 0.4pt}}}$}}
\newcommand{\proof}[1]{{\bf Proof.} #1~$\qed$.}

This invited contribution summarizes the talk given by the author 
at the conference 
``Secondary Calculus and Cohomological Physics, August 24--31, 1997, 
Moscow, Russia''. The first part is based on work done in 
collaboration with R.~Fulp, T.~Lada and J.~Stasheff \cite{BFLS}. The
second part is based on work done in collaboration with M.~Henneaux
\cite{BaHe}. 

\section{Sh Lie structure of brackets on the horizontal complex}

\subsection{The horizontal complex as a resolution for local functionals}

In the approach of Gel'fand, Dickey and Dorfman to functionals in
field theory \cite{GeDi1,GeDi2,GeDo1,GeDo2,GeDo3} (see \cite{Dic1} for a
review), one replaces local functionals satisfying appropriate boundary
conditions by equivalence classes of local functions. 

Let $M$ be an $n$-dimensional manifold homeomorphic
to ${bf {R}}^n$ with coordinates denoted by $x^i$ and 
$\pi:E=M\times V\rightarrow M$ a trivial vector
bundle of fiber dimension $k$ over $M$. 
The coordinates of $V$ are denoted by $u^a$.

Let $J^{\infty}E$ denote the
infinite jet bundle of $E$ over $M$ with $\pi^{\infty}_E:J^{\infty}E
\rightarrow E$ and $\pi^{\infty}_M:J^{\infty}E \rightarrow M$ the
canonical projections.
The vector space of smooth sections of $E$ with
compact support will be denoted $\Gamma E$.
For each section
$\phi$ of $E$, let $j^{\infty}\phi$ denote the induced section
of the infinite jet bundle $J^{\infty}E$.
The bundle
\begin{eqnarray}
\pi^\infty : J^\infty E=M\times V^\infty \rightarrow M   
\end{eqnarray}
then has induced coordinates given by
\begin{eqnarray}
(x^i,u^a,u^a_i,u^a_{i_1i_2},\dots,).
\end{eqnarray}

\begin{definition}
A local function on $J^{\infty}E$ is the pullback of a smooth
function on some finite jet bundle $J^pE$, i.e., a composite $J^\infty
E \to J^pE \to {bf {R}}$. In local coordinates, a local function
$L(x,u^{(p)})$ is a smooth function in the coordinates $x^i$ and the
coordinates $u^a_I$, where the order $|I| = r$ of the multi-index $I$
is less than or equal to some integer $p$.
The space of local functions will be denoted by $\CJ$
\end{definition}

Let $\nu$ denote a fixed volume form on  $M$ and let $\nu$ also denote
its pullback $(\pi^{\infty}_E)^*(\nu)$ to $J^{\infty}E$.
In coordinates, $\nu = d^nx =
dx^1\wedge\cdots\wedge dx^n$.

\begin{definition}
A local functional
\begin{eqnarray}
{\mathcal L}[\phi]=\int_M L(x,\phi^{(p)}(x)) \nu = \int_M (j^\infty
\phi)^*  L(x,u^{(p)}) \nu
\end{eqnarray}
is the integral over $M$ of a local function evaluated for sections
$\phi$ of $E$ of compact support.
The space of local functionals ${\mathcal F}$ is the vector space of
equivalence classes of local functionals, where two local functionals
are  equivalent if they agree for all sections of compact support.
\end{definition}
The total derivative $D_i$ along $x^i$ is
\begin{eqnarray}
D_i=\frac {\partial}{\partial x^i}+u^a_{iJ}\frac {\partial}{\partial
u^a_J}.
\end{eqnarray}
The horizontal complex $(\Omega^*,d_H)$ is the exterior algebra in the
$dx^i$ with coefficients that are local functions and differential
$d_H=dx^iD_i$.

\begin{lemma}\label{l4}
The vector space of local functionals ${\mathcal F}$ is isomorphic to the
cohomology group $H^n(d_H)$.
\end{lemma}

\proof{
A complete proof can be found for instance in \cite{Olv}.
}

Furthermore, one can also show (see for instance \cite{Olv}) 
that the horizontal complex without the
constants provides a homological resolution of ${\mathcal F}$, i.e., the
complex\\
$\begin{array}{ccccccccc}\\ & d_H & & d_H & & d_H & & d_H&\\ 
    \Omega^{0}/{bf{R}}&\longrightarrow &\dots &
\longrightarrow &
\Omega^{n-2} &\stackrel {\textstyle}{\longrightarrow}
&\Omega^{n-1}&\stackrel {\textstyle}{\longrightarrow}
&\Omega^{n}\\ \downarrow\eta & & & &
\downarrow\eta & & \downarrow\eta & & \downarrow\eta\\ 
0 &\longrightarrow &\dots & \longrightarrow &
0&\longrightarrow & 0 &\longrightarrow & H_0={\mathcal F}
\end{array}$\\
\vskip1ex
\noindent is exact.

\subsection{Algebraic definition of brackets}

In the simplest case, brackets in field theory are defined by
splitting the coordinates $u^a$ of an even-dimensional $V$ in two sets
$u^\alpha$ and $\pi_\beta$ and declaring these two
be canonically conjugate, i.e.,
$\{u^\alpha(x),\pi_\beta(y)\}=\delta^\alpha_\beta\delta(x,y)$, while
all other brackets vanish. Brackets of local functionals are then
defined in terms of functional derivatives by 
\be
\{{\mathcal L}_1,{\mathcal L}_2\}=
\int_M \frac{\delta {\mathcal L}_1}{\delta u^\alpha(x)}
\frac{\delta {\mathcal L}_2}{\delta \pi_\beta(x)} 
\nu-(1\longleftrightarrow 2).
\ee
In the case of sections with compact support, the functional
derivative  is equal to
the Euler-Lagrange derivative pulled back to the corresponding
section: 
\be
\frac{\delta {\mathcal L}_1}{\delta u^\alpha(x)}=(j^\infty
\phi)^*\frac{\delta L_1}{ \delta u^\alpha},
\ee
the Euler-Lagrange operator being defined by 
\begin{eqnarray}
\frac{\delta L_1}{\delta u^a}
=\frac{\partial L_1}{\partial u^a}-\partial_i\frac{\partial
L_1}{\partial u^a_i}+\partial_i\partial_j\frac{\partial L_1}{\partial
u^a_{ij}}-...=(-D)_I(\frac{\partial L_1}{\partial u^a_I}).
\ee
This suggest taking
\be
\{L_1\nu,L_2\nu\}=
[\frac{\delta {L}_1}{\delta u^\alpha}
\frac{\delta {L}_2}{\delta \pi_\beta} -(1\longleftrightarrow
2)]\nu
\ee
as a definiton of a bracket on $\Omega^{n}$. 
Since Euler-Lagrange derivatives annihilate total divergences, we have
\be
\{d_H R,L_2\nu\}=0
\ee
for $R\in \Omega^{n-1}$, so that there is indeed a well defined
induced bracket on ${\mathcal F}\simeq H^n(d_H)$, given by  
\be
\{\cdot,\cdot\}:{\mathcal F}\wedge{\mathcal F}\longrightarrow
{\mathcal F}, \nonumber\\
\{[L_1\nu],[L_2\nu]\}=[\{L_1\nu,L_2\nu\}].
\ee

As in the case of standard mechanics, 
brackets need not be defined in terms of
Darboux coordinates as above. Let $TDO(E)$ be the space of total
differential operators. In coordinates, a total differential
operator is given by an operator of the form $Z^ID_I$, with $Z^I\in
Loc(E)$. Let $\omega^{ab}$ be a square matrix of total differential 
operators, $\omega^{ab}=\omega^{abI}D_I$. We then have the following
general definition (see e.g. \cite{Olv,Dic1})~:

\begin{definition}
The operator matrix $\omega^{ab}\in
TDO(E)$ defines a skew-symmetric bracket on $\Omega^{n}$ through
\be
\{\cdot,\cdot\}:\Omega^{n}\wedge\Omega^{n}
\longrightarrow\Omega^{n},\nonumber\\
\{L_1\nu,L_2\nu\}=\frac{1}{2}[\omega^{ab}(\frac{\delta L_1}{ \delta
  u^a})\frac{\delta L_2}{ \delta u^b}-(1\longleftrightarrow 2)]
\nu.\label{3}
\ee
\end{definition}

In principle, one could allow for an arbitrary combination of 
the Euler-Lagrange derivatives,
$\omega^{abIJ}D_I\frac{\delta L_1}{ \delta u^a}
D_J\frac{\delta L_2}{ \delta u^b}$, with $\omega^{abIJ}\in Loc(E)$.
Since we are only interested in the brackets induced in
${\mathcal F}$, this case can be reduced to the above 
by integrations by parts giving $d_H$ exact
terms projecting to zero in ${\mathcal F}$. More precisely, one can show
\cite{Olv} that a skew-symmetric bracket on ${\mathcal F}$ is uniquely
determined by a functional two vector which in turn is uniquely
determined by a skew-adjoint matrix of total differential operators
$\omega^{ab}$~: 
$\omega^{abI}D_If_a=-(-D)_I[\omega^{baI}f]$ for all $f\in Loc(E)$.

\subsection{Jacobi identity}

The requirement that the induced bracket on ${\mathcal F}$ satisfies the
Jacobi identity is equivalent to 
\be
\{\{L_1\nu,L_2\nu\},L_3\nu\}+cyclic=d_H R(L_1\nu,L_2\nu,L_3\nu)\label{j}
\ee
for all $L_1\nu,L_2\nu,L_3\nu\in \Omega^n$ with 
$R(L_1\nu,L_2\nu,L_3\nu)\in
\Omega^{n-1}$. This imposes restrictions 
on the skew-adjoint total differential operator $\omega^{ab}$. If
these restrictions are satisfied, one says \cite{Olv,Dic1} that  
$\omega^{ab}$ is
Hamiltonian. The simplest examples of Hamiltonian $\omega^{ab}$'s are
given by the constant symplectic matrix $\sigma^{ab}$ considered above
in the case of the bracket in Darboux coordinates, as well as the
total differential operator $D_x$ of the KDV equation, where the
dimension $n$ of the base space and $k$ of the fiber are both equal to
$1$ (see also \cite{Dic2}).

We thus see that in the case where the induced bracket is a Lie bracket,
the bracket $l_2\equiv\{\cdot,\cdot\}$ on $\Omega^{n}$ induces a
completely skew-symmetric bracket with three entries defined by
\be
l_3:\wedge^3\Omega^{n}\longrightarrow\Omega^{n-1}\nonumber\\
l_3(L_1\nu,L_2\nu,L_3\nu)=R(L_1\nu,L_2\nu,L_3\nu).
\ee

\subsection{Sh Lie algebras from homological resolutions of Lie
  algebras}

The algebraic situation we have is the following. Let ${\mathcal F}$ be a
vector space and $(X_*,l_1)$ be a homological resolution thereof, i.e.,
we have a graded vector space
$X_*=X_0\oplus X_1\oplus\dots,$ with a differential
$l_1:X_k\longrightarrow X_{k-1}$, such that $H_0(l_1)\simeq {\mathcal F}$
and $H_k(l_1)=0, k>0$. 

(In our case above, ${\mathcal F}$ is the space of local functionals, 
$X_*$ is the horizontal complex without the constants, if we define 
the grading to be $n$ minus the horizontal form degree and identify 
$l_1=d_H$.)

Let $[\cdot,\cdot]$ be a homomorphism from $\wedge^2{\mathcal F}$ to 
${\mathcal
  F}$ and $l_2$ a homomorphism from  $\wedge^2 X_0$ to $X_0$ such that
  $[[x_1],[x_2]]=[l_2(x_1,x_2)].$
\begin{lemma}
The bracket $[\cdot,\cdot]$ is well defined and a Lie bracket iff 
$\forall\ x_1,x_2,x_3\in X_0,$ $\forall\ y_1\in X_1$, 
$\exists\ y_2,y_3\in X_1$
such that 
\be
l_2(l_1 y_1,x_1)=l_1 y_2\label{1}\\
l_2(l_2(x_{1},x_{2}),x_{3})+cyclic=l_1 y_3.\label{2}
\ee
\end{lemma}
\proof{
The proof is straightforward and can be found in \cite{BFLS}.
}

Let $sX_*$ be the graded vector space defined by $(sX)_k=X_{k-1}$.
Let $T^csX_*=\oplus_{k=0}sX_*^{\otimes^k}$ be the tensor coalgebra of 
$sX_*$
with standard diagonal 
\be
\Delta (sx_1\otimes\dots\otimes sx_k)
=\Sigma_{j=0}^k(sx_1\otimes\dots\otimes sx_j)\otimes(sx_{j+1}\dots\otimes
sx_k). 
\ee
Let $\bigwedge sX_*$ be the graded
symmetric subcoalgebra of $T^cX_*$. Any homomorphism $f$ from 
$\bigwedge^k sX_*$ to $sX_*$ can be extended in a unique way 
as a coderivation
to $\bigwedge sX_*$ (see for instance \cite{LaSt,LaMa,Kje} for more 
details).
Let us suppose this has been done in particular for $l_1$ and let us
denote the resulting coderivation by $\hat l_1$.
Let us
define the total degree $tot$ or (just degree for short) 
in $T^csX_*$ of the element 
$sx_1\wedge\dots\wedge sx_r$ to be 
$\Sigma_{i=1}^r|sx_i|=\Sigma_{i=1}^r|x_i|+r$, while the 
resolution degree $res$ of this element is defined to be
$tot-r=\Sigma_{i=1}^r|x_i|$. These definitions imply that $\hat l_1$ is
of resolution degree and total degree $-1$.
Let $[\cdot,\cdot]$ be the graded commutator of coderivations with
respect to the total degree.
\begin{lemma}\label{l1}
The map $l_2$ 
can be extended as a degree $-1$ coderivation $\hat l_2$ of resolution
degree $0$ on
$\bigwedge sX_*$ in such a way that $[\hat l_1,\hat l_2]=0$.
\end{lemma}
\proof{
It is enough to verify the statement for $sx_1\wedge sx_2$. For
$x_1,x_2\in X_0$, i.e, $sx_1\wedge sx_2$ of resolution degree $0$,
the lemma is true
since $\hat l_1$ vanishes on elements of resolution degree $0$ and
$res(l_2(sx_1\wedge sx_2))=0$. For
$res(sx_1\wedge sx_2)=1$, we can assume without loss of generality
that $res(sx_1)=1$, $res(sx_2)=0$. 
We have $l_2\hat l_1(sx_1\wedge
sx_2)=l_2(l_1sx_1,sx_2)+(-)^{|sx_1|}l_2(sx_1,l_1sx_2)
=l_2(l_1sx_1,sx_2)=-l_1sy$
because of (\ref{1}). Hence, we can achieve $[\hat l_1,\hat l_2]=0$ in
resolution degree 1 if we define $l_2(sx_1\wedge sx_2)=sy$.
Suppose the lemma
is true for $res(sx_1\wedge sx_2)=k\geq 1$. For $res(sx_1\wedge
sx_2)=k+1$, we have $0=l_2\hat l_1(\hat l_1(sx_1\wedge
sx_2))=-l_1l_2(\hat l_1(sx_1\wedge sx_2))$. 
Since $res(l_2(\hat l_1(sx_1\wedge
sx_2)))>0$, acyclicity of $l_1$ implies that there exists $sy$ such 
$l_2(\hat l_1(sx_1\wedge sx_2))=-l_1(sy)$. Hence we can define
$l_2(sx_1\wedge sx_2)=sy$ in resolution degree $k+1$.
}
\begin{lemma}\label{l2}
There exists a degree $-1$ coderivation $\hat l_3$ of resolution degree
$1$ on $\bigwedge sX_*$ such that $\hat l_1\hat l_3+\hat l_2\hat
l_2+\hat l_3\hat l_1=0$.
\end{lemma}
\proof{
It is enough to verify the relation for $sx_1\wedge sx_2\wedge
sx_3$. In resolution degree $0$, $\hat l_2\hat l_2(sx_1\wedge sx_2\wedge
sx_3)$ is another expression for the left hand side of the Jacobi
identity for the bracket $l_2$. Using (\ref{2}), there exists $l_3$
such that $(l_1l_3+\hat l_2\hat l_2)(sx_1\wedge sx_2\wedge
sx_3)=0$. Since we are in resolution degree $0$, this can also be
written as $(l_1l_3+\hat l_2\hat l_2+l_3\hat l_1)(sx_1\wedge sx_2\wedge
sx_3)$. Let us suppose that the relation holds for 
$sx_1\wedge sx_2\wedge sx_3$ of resolution degree
$k\geq 1$. If $sx_1\wedge sx_2\wedge sx_3$ is of resolution degree $k+1$,
then $\hat l_1(sx_1\wedge sx_2\wedge
sx_3) $ is of resolution degree $k$ and hence 
$0=(l_1l_3+\hat l_2\hat l_2+l_3\hat l_1)\hat l_1(sx_1\wedge sx_2\wedge
sx_3)=l_1(l_3\hat l_1+\hat l_2\hat l_2)(sx_1\wedge sx_2\wedge
sx_3)$. Acyclicity of $l_1$ in resolution degree $>0$ then implies that
one can extend $l_3$ to resolution degree $k+1$ in such a way that 
$\hat l_1\hat l_3+\hat l_2\hat
l_2+\hat l_3\hat l_1=0$ holds.
}
\begin{lemma}
Acyclicity of $l_1$ in resolution degree $>0$, i.e.,
$l_1(sx)=0\Longleftrightarrow sx=l_1sy$ for $res(sx)>0$,
implies acyclicity of $\hat l_1$ in the space of coderivations of
resolution degree $>0$~: for any $\hat d$ with $res(\hat d)=k>0$ such
that $[\hat l_1, \hat d]=0$, 
there exists $\hat t$ with  $res(\hat t)=k+1$ such that 
$\hat d=[\hat l_1, \hat t]$.
\end{lemma}
\proof{
Let $\rho$ be a contracting homotopy for $l_1$, i.e., $l_1\rho+\rho
l_1=Id$ when applied to elements of resolution degree $>0$. Without
loss of generality, we can assume that $\hat d$ is the extension as a
coderivation of a homomorphism $d: sX_*^{\bigwedge^k}$ to $sX_*$. 
Hence, we
have on $sX_*^{\bigwedge^k}$, $d=(l_1\rho+\rho l_1)d$. 
The cocycle condition then implies
that $d=l_1\rho d-\rho d \hat l_1=[\hat l_1,\rho d]$.
}
\begin{theorem}\label{t1}
There exist degree $-1$ coderivations $\hat l_k$ of resolution degrees
$k-2$ for $k\geq 4$ such that the degree $-1$ coderivation
$s=\hat l_1+\hat l_2+\hat l_3+\hat l_4+\dots$ is a differential. 
\end{theorem}
\proof{
The condition $\frac{1}{ 2}[s,s]=0$ at resolution degree $k-1$ gives
$I_k=\hat l_1\hat l_{k}+\hat l_2\hat l_{k-1}+\dots+\hat l_k\hat l_1=0$.
We have shown above that this relation holds for $k\leq 3$. Suppose
that it holds for all $l\leq k$, with $k\geq 3$. 
The graded Jacobi identity for
$[\cdot,\cdot]$ implies that $\frac{1}{ 2}[s,[s,s]]=0$. At resolution
degree $k$, this identity reads $[\hat l_1,I_{k+1}]+[\hat
l_2,I_{k}]+\dots+ [\hat l_{k+1},I_1]=0$. The recursion hypothesis then
implies that $[\hat l_1,\hat l_1\hat l_{k+1}+\hat l_2\hat l_k+\dots
\hat l_k \hat l_2+\hat l_{k+1}\hat l_1]=0$. The first term cancels with
the last one, so that $[\hat l_1,\hat l_2\hat l_k+\dots
\hat l_k \hat l_2]=0$. Since $k\geq 3$, the resolution degree of 
$\hat l_2\hat l_k+\dots
\hat l_k \hat l_2$ is $\geq 1$ so that the previous lemma implies the
existence of $\hat l_{k+1}$ such that $\hat l_2\hat l_k+\dots
\hat l_k \hat l_2=-[\hat l_1,\hat l_{k+1}].$
}
The associated skew
coderivations $\tilde l_k$ 
on the skew coalgebra $\bigwedge_s X_*$, 
(see \cite{LaSt,LaMa,Kje} for details) form an sh Lie algebra.

\subsection{Reduced form}

In the construction above, only $l_1$ and $l_2$ were initially fixed on
$X_0$. In the extension of $l_2$ to $sX_*^{\bigwedge^2}$, and the
construction of $l_3, l_4, ...$, we have the liberty, at each stage
where we made a choice, to add $l_1$ exact terms.
We will consider sh Lie algebras
constructed in this way and corresponding to different choices to 
be equivalent.

In the case of local functionals, the resolution is trivial for
resolution degree $>n$, which means horizontal form degree $<0$. This
means that the construction yields non vanishing brackets at most up to
the bracket $l_{n+2}$. 

Furthermore, the bracket $l_2$ on $X_0$ is such that (\ref{1}) holds
with $0$ on the right hand side. This is because the bracket is
expressed in terms of Euler-Lagrange derivatives which annihilate
$d_H$ exact $n$-forms. The following is due to M. Markl \cite{Mar}:
\begin{theorem}
If (\ref{1}) holds with $0$ on the right hand side, the extension of 
$l_2,l_3$
to $X_*$ can be taken to be trivial. The sh Lie algebra
is equivalent to an sh Lie algebra where only the brackets
$l_1,l_2,l_3$ are non vanishing, i.e., $s=\hat l_1+\hat l_2+\hat l_3$
is a differential.
\end{theorem} 
\proof{
In the proof of lemma \ref{l1}, we can take $sy=0$ so that we can
define $l_2(sx_1\wedge sx_2)=0$ if the resolution degree of
$sx_1\wedge sx_2$ is $1$ and satisfy $[\hat l_1,\hat
l_2]=0$,
i.e., $\hat l_1\hat
l_2=0$ and $\hat l_2\hat l_1=0$. If we extend $l_2$ to be zero in all
resolution degrees $>1$, these two relations continue to hold on all
of $sX_*^{\bigwedge^2}$. We have through (\ref{2}) that there exist
$l_3$ on $sX_*^{\bigwedge^3}$ such that 
$l_1l_3+\hat l_2\hat l_2=0$ on elements in resolution degree $0$.
We can choose $l_3$ to be zero on $\hat l_1$-exact
terms in resolution degree $0$ and in resolution degree $>0$,
so that $\hat l_1\hat l_3+\hat l_2\hat l_2+\hat l_3\hat
l_1=0$ is non trivial only in resolution degree $0$ where it reduces to
$l_1l_3+\hat l_2\hat l_2=0$. In resolution degree $3$, we have to
consider the expression $\hat l_3\hat l_2+\hat l_2\hat l_3$ which
reduces to $\hat l_3\hat l_2$ because $\hat l_2$ vanishes on elements
of resolution degree $>0$. Let $\rho$ be a contracting homotopy for
$l_1$. We have $l_3=(l_1\rho+\rho l_1)l_3=l_1\rho l_3-\rho
l_2\hat l_2$. Hence, we can choose $l_3=-\rho l_2\hat l_2$.
The identity $[\hat l_1,\hat l_3\hat l_2+\hat l_2\hat l_3]=0$ reduces
to $0=\hat l_1\hat l_3\hat l_2=\hat l_2\hat l_2\hat l_2$ because of
our assumptions on $\hat l_2,\hat l_3$, so that $l_3\hat l_2=-\rho
l_2\hat l_2\hat l_2=0$. We then can take $\hat l_4$ and all
higher order coderivations to be zero and $s=\hat l_1+\hat
l_2+\hat l_3$ is a differential.
}

\subsection{Sh Poisson algebra ?}

In the above considerations, we have only been concerned with the Lie
algebra aspects of Poisson brackets in field theory. The Poisson
brackets in mechanics are in addition derivations with respect to the
product of functions. The problem for Poisson brackets in field theory
is that there is no product for local functionals. There is however a
well defined product on $\Omega^n$, but it does not induce a product
on ${\mathcal F}$. In coordinates, it is defined by 
$L_1\nu\cdot L_2\nu=L_1L_2\nu$. This product can be expressed
in an invariant way on a Riemannian manifold in terms of the Hodge
star and the wedge product. The brackets (\ref{3}) satisfy
the Leibnitz rule up to homotopy. Indeed, the
skew adjointness of the operator $\omega^{ab}$ implies that 
\be
\{L_1\nu,L_2\nu\cdot L_3\nu\}=\omega^{ab}(\frac{\delta L_1}{ \delta
  u^a})\frac{\delta L_2L_3}{\delta u^b}\nu+d_HR_1\nonumber\\
=D_I[\omega^{ab}(\frac{\delta L_1}{ \delta u^a})]
\frac{\partial (L_2L_3)}{ \partial u^b_I}\nu +d_HR_2\nonumber\\
=\{L_1\nu,L_2\nu\}\cdot L_3\nu+
\{L_1\nu,L_3\nu\}\cdot L_2\nu +d_HR_3,
\ee
for some $R_i\in \Omega^{n-1}$ obtained by integrations by parts.
This suggests that in terms of representatives, i.e., elements of
$\Omega^n$, there should not only be a sh Lie structure, but
rather an appropriately defined sh Poisson structure. 

A different approach would be the following.
The properties of a Poisson bracket in mechanics,
including the Leibnitz rule with respect to the product,
can be summarized by defining the bracket in terms of a two vector
whose Schouten bracket with itself vanishes. In field theory, the
situation is very similar, the Poisson bracket as defined in
(\ref{3}) is in fact associated with a functional two vector. Indeed,
as mentioned before, 
one can show \cite{Olv} that a functional two vector is uniquely
determined by a skew adjoint operator $\omega^{ab}$.
Because the Schouten bracket for multi-vector fields together with
the ordinary product for functions define a Gerstenhaber algebra, we
can expect an sh Gerstenhaber algebra, which have been used recently in
different contexts \cite{LiZu,GeVo,KVZ}, to be useful here as well.

\section{Batalin-Vilkovisky bracket and Poisson bracket}

\subsection{The local antibracket in cohomology}

In the Batalin-Vilkovisky formalism, we consider instead of the
exterior algebra over $dx^i$ with coefficients belonging to $Loc(E)$,
the space $Loc(E)\otimes\Lambda(C^\alpha_I,u^*_{aI},
C^*_{\alpha I},dx^i)$, where
$C^\alpha_I,u^*_{aI},C^*_{\alpha I}$ are the jet-bundle analogs of the
ghosts, the antifields, the antighosts and their derivatives. The
original fields $u^a_I$ and the coordinates $x^i$ with their
differentials $dx^i$ are defined to be of ghost number $0$, while
$C^\alpha_I,u^*_{aI},C^*_{\alpha I}$ are respectively of ghost number
$1$, $-1$ and $-2$. If we define the
fields $\phi^A\equiv(u^a_I,C^\alpha_I)$, the
total derivative is extended to the new generators and reads:
\be
D_i=\frac {\partial}{\partial x^i}+\phi^A_{Ii}\frac{\partial}
{\partial \phi^A_I}+
{\phi^*_A}_{Ii}\frac{\partial}{\partial {\phi^*_A}_I}.
\ee
The BRST differential is denoted by $s$~; it is of ghost number $1$ 
and satisfies $[s,d_H]=0$, so
that we have the bicomplex $(\Omega^{*,*},s,d_H)$. The 
graded commutator of graded derivations involves the total
degree which is the ghost number plus the horizontal form degree.

The bracket is the BV antibracket, adapted from its functional
expression to the $n$-forms as in the first part~: it is given by 
\be
\{L_1\nu,L_2\nu\}=
[\frac{\delta^R {L}_1}{ \delta \phi^A}
\frac{\delta^L {L}_2}{ \delta \phi^*_A} 
-\frac{\delta^R {L}_1}{ \delta \phi^*_A}
\frac{\delta^L {L}_2}{ \delta \phi^A}]\nu.\label{4}
\ee
It satisfies a graded version of antisymmetry and Jacobi identity up
to exact terms because it is expressed in Darboux coordinates so that
the bracket induced in $H^{*,n}(d_H)$ is a graded Lie bracket
corresponding to the usual Batalin-Vilkovisky bracket for local
functionals. 

The important property of the BRST differental $s$ is that it is
canonically generated through a variant of the local antibracket 
by a local functional $S=[L\nu]$ which is a solution of
the BV master equation 
\be
\{S,S\}=0\Longleftrightarrow \{L\nu,L\nu\}=d_HR
\ee
 for $R\in \Omega^{*,n-1}$,
\be
s=\{L\nu,\cdot\}_{alt}=D_I(\frac{\delta^R {L}}{\delta \phi^A})
\frac{\partial^L }{ \partial \phi^*_{AI}}-
D_I(\frac{\delta^R {L}}{\delta \phi^*_A})
\frac{\partial^L }{ \partial \phi^A_{I}}.
\ee

A proper solution of the master equation, in the case of an
irreducible gauge theory, is of the form 
\be
S=\int_M\ (L_0+u^*_aR^{aI}_\alpha D_IC^\alpha+...)\nu,
\ee
where $L_0$ is the starting point Lagrangian, $R^{aI}_\alpha D_I$
describe a generating set of gauge symmetries and the higher terms are
determined by requiring that $S$ is a solution to the master equation.

The local antibracket (\ref{4}) also induces a well
defined bracket in 
$H^{*}(s,{\mathcal F})\simeq H^{*}(s,H^n(d_H))\simeq H^{*,n}(s|d_H)$.
More precisely,
\begin{eqnarray}
\{\cdot,\cdot\} : H^{g_1,n}(s|d_H)\times H^{g_2,n}(s|d_H)\longrightarrow 
H^{g_1+g_2+1,n}(s|d_H)\nonumber\\
\{[L_1\nu],[L_2\nu]\}=[\{L_1\nu, L_2\nu\}]\label{abm}.
\end{eqnarray}
The
subspace $H^{-1,n}(s|d_H)$ equipped with the antibracket defines a
subalgebra of $H^{*,n}(s|d)$ which we denote by ${\mathcal S}$,
\begin{eqnarray}
{\mathcal S}=(H^{-1,n}(s|d),\{\cdot,\cdot\}).
\end{eqnarray}

The BRST cohomology groups contain the physical information of the problem.
It is this last bracket, induced by the local antibracket in the local
BRST cohomology groups $H^{*,n}(s|d_H)$ that we will relate to the
Poisson bracket. It has been shown in \cite{BBH}, that the cohomology
group in negative ghost numbers $g$ describe the characteristic cohomology
of the problem, i.e., the $n+g$ forms which are closed (under $d_H$) on the
stationary surface $\Sigma$, which is 
defined by $D_I\frac{\delta L_0}{\delta u^a}=0$ in the jet-bundle, 
without being exact on this surface. In
positive ghost number, the cohomology of $s$ describes equivariant
cohomology associated to the gauge transformations on the stationary
surface. The induced antibracket provides these cohomology groups with a
graded Lie algebra structure. One can relate this Lie algebra
structure in general with the graded Lie algebra induced by the
Batalin-Fradkin-Vilkovisky extended Poisson bracket induced in the
local Hamiltonian BRST cohomology groups \cite{BaHe}. Since this is
rather technical we will consider below only the more transparent 
cases of ordinary
mechanics and non degenerate field theories.

\subsection{Ordinary mechanics}

The simplest case is ordinary mechanics in Hamiltonian form. The first
order action 
\be 
S=\int dt\ \dot qp -H
\ee
is by itself a proper solution to the master equation 
since there are no gauge invariances. The BRST differential is given
by 
\be
s=(D_t)^k(\dot q-\frac{\partial H}{\partial p})
\frac{\partial }{ \partial p^{*(k)}}+
(D_t)^k(-\dot p-\frac{\partial H}{\partial q})
\frac{\partial}{ \partial q^{*(k)}},
\ee
where $p^{*(k)},q^{*(k)}$ denote the $k$-th derivatives of the
antifields $p^*,q^*$ with
respect to time $t$. We have to compute $H^{*,1}(s|d_H)$. This group
is computed by the descent equations technique, so that we must first
compute $H^{*}(s)$. It is straightforward to see that the time
derivatives of $p,q$ of order at least equal to $1$ 
and the antifields with all their derivatives can be eliminated form
the cohomology which is equal to $C^\infty(p,q)$. Let $f\in
C^\infty(p,q)$. We have to solve $sg dt +d_H f=0$ in ghost number
$-1$. 
This equation
implies that $g=dt [\frac{\partial f}{\partial q}p^*-
\frac{\partial f}{\partial p}q^*]$ with $f$ satisfying $[f,H]_P=0$,
where $[\cdot,\cdot]_P$ is the standard Poisson bracket. In
other words, the bottom $f$ of the descent equations is a first
integral, while the corresponding top in ghost number $-1$
is the associated Hamiltonian
vector field upon identification of $q^*,p^*$ with
$\frac{\partial}{\partial q},\frac{\partial}{\partial p}$.
This identification implies that the Lie algebra ${\mathcal S}$
is isomorphic to the Lie algebra of Hamiltonian vector fields for
first integrals equipped with the standard Lie bracket for vector
fields:
\be
\{\frac{\partial f_1}{\partial q}p^*-\frac{\partial f_1}{\partial p}q^*,
\frac{\partial f_2}{\partial q}p^*-\frac{\partial f_2}{\partial
  p}q^*\}=
\frac{\partial [f_1,f_2]_P}{\partial q}p^*-\frac{\partial
  [f_1,f_2]_P}
{\partial p}q^*.
\ee
This last algebra is isomorphic to the Poisson algebra of first
integrals, which gives the desired relation between the local
antibracket induced in cohomology and the Poisson bracket.

\subsection{Non degenerate field theories}

In a field theory without gauge symmetries, $S=\int L_0\nu$ is a proper
solution to the master equation and the BRST symmetry reduces to 
\be
s=D_I(\frac{\delta^R {L_0}}{ \delta u^a})
\frac{\partial^L }{ \partial u^*_{aI}}\equiv\delta,
\ee
which is called the Koszul(-Tate) differential associated to the
stationary surface $\Sigma$. Its cohomology in the algebra
$Loc(E)\otimes\Lambda(u^*_{aI},dx^i)$ is given by
$H^{0}(\delta)\simeq Loc(E)/I\otimes\Lambda(dx^i)$, where $I$ is the
ideal of horizontal forms vanishing on $\Sigma$, while
$H^{k}(\delta)=0$ 
for
$k<0$. We are interested in $H^{-1,n}(\delta|d_H)$ in the algebra 
$Loc(E)/{bf R}\otimes\Lambda(u^*_{aI},dx^i)$.
\begin{theorem}[Cohomological formulation of Noether's first theorem]

\be
H^{-1,n}(\delta|d_H)\simeq H^{n-1,0}(d_H|\delta).
\ee
\end{theorem}
\proof{
A cocycle in the first group satisfies $\delta a +d_H j=0$. If we
consider a different representative $a^\prime=a+\delta()+d_H()$, the
corresponding $j$ is modified only by $\delta$ and $d_H$ exact
terms. Hence, the map $f([a])=[j]$ from $H^{-1,n}(\delta|d_H)$ to
$H^{0,n-1}(d_H|\delta)$ is well defined. It is injectif because if
$[j]=0$, i.e., $j=d_H()+\delta()$, then acyclicity of $\delta$ in
ghost number $-1$ implies that $a=\delta() +d_H()$. Inverting the role
of $\delta$ and $d_H$ and using the acyclicity of $d_H$ in form degree
$n-1$, we find that the map $f$ is bijective.
}
The interpretation of $H^{n-1,0}(d_H|\delta)$ is clear. Since $\delta$
exact forms in ghost number $0$ are precisely given by forms which
vanish on $\Sigma$, this group describes the characteristic cohomology
in form degree $n-1$, or in dual language, the equivalence classes of
currents that are conserved on the stationary surface, $D_ij^i=0$ on
$\Sigma$, where two currents are equivalent if they differ by a
current of the form $D_kS^{[ki]}$ on $\Sigma$, with $j^i,S^{[ki]}\in
Loc(E)$. 

By allowed redefinitions, (adding $d_H$ exact terms, or in dual
notation, doing integrations by parts), 
we can assume that a representative of an
element of $H^{-1,n}(\delta|d_H)$ is of the form $a=u^*_aX^a\nu$, with
$X^a\in Loc(E)$.
The cocycle condition reads 
\be
\frac{\delta L_0}{\delta u^a}X^a+D_ij^i=0,
\ee
for some $j^i\in Loc(E)$. Hence, the evolutionary vector field
$X^a\frac{\partial}{\partial u^a}$ defines a variational symmetry of
$L_0$. The coboundary conditions implies that a trivial variational 
symmetry is of
the form $X^au^*_a=\delta(\frac{1}{2}u^*_{bJ}u^*_{aI}\mu^{[aIbJ]})+D_i
k^i$, for some $\mu^{[aIbJ]}\in Loc(E)$ which are antisymmetric under
  the exchange $aI \longleftrightarrow bJ$ and some $k^i$ which are
  linear polynomials in $u^*_{aI}$. Taking Euler-Lagrange derivatives
  with respect to $u^*_a$, this implies $X^a=(-D)_I[D_J\frac{\delta
    L_0}{\delta u^b}\mu^{[aIbJ]}]$. In particular, $X^a$ must vanish
  on the stationary surface. 

Hence, the lemma above expressed that
  there is an isomorphism between equivalence classes of variational
  symmetries (two variational symmetries are equivalent if they differ
  by a trivial symmetry defined above) and equivalence classes of
  conserved currents. This is precisely the content of Noether's first
  theorem. 

If we identify $u^*_a$ with $\frac{\partial}{\partial u^a}$, 
the antibracket $\{[a_1],[a_2]\}$ can be identified with the ordinary
Lie bracket for variational symmetries induced by the Lie bracket for
evolutionary vector fields,
\be
[\{a_1,a_2\}]=[(\frac{\delta (u^*_aX^a_1)}{\delta u^b}X^b_2-
\frac{\delta (u^*_aX^a_2)}{\delta u^b}X^b_1)\nu=[u^*_a[X_2,X_1]^a\nu],
\ee
so that ${\mathcal S}$ is isomorphic to the Lie algebra of variational
symmetries equipped with the bracket for evolutionary vector fields.

{}From the isomorphism, we know that there is an induced bracket in
$H^{n-1,0}(d_H|\delta)$. An explicit calculation involving the
properties of Euler-Lagrange derivatives gives \cite{BaHe}
\be
\{[j_1],[j_2]\}=-[D_I(X^a_1)\frac{\partial}{\partial
  u^a_I}j_2]=[D_I(X^a_2)\frac{\partial}{\partial u^a_I}j_1]. 
\ee
This bracket has been proposed by Dickey \cite{Dic1} as a covariant
way to define Poisson brackets for conserved currents, so that
${\mathcal S}$ is isomorphic to the Lie algebra of (equivalence classes
of) conserved currents equipped with the Dickey bracket. Finally, 
one can show
\cite{Dic1,BaHe} that this Lie algebra is equivalent to the Lie
algebra of conserved charges $Q=\int dx^1\wedge\dots\wedge dx^{n-1}
j^0$ equipped with the local Poisson bracket induced in the space of
functionals on space alone. This gives the desired relation between
the antibracket and the Poisson bracket.

\section*{Acknowledgments}
The author wants to thank Professor H. Kleinert for hospitality in his
group. 

\bibliographystyle{amsalpha}

\begin{thebibliography}{A}

\bibitem[BFLS]{BFLS} G. Barnich, R. Fulp, T. Lada and J. Stasheff,
  {\em The sh Lie structure of Poisson brackets in Field Theory}, to
  appear in Commun. Math. Phys.

\bibitem[BaHe]{BaHe} G. Barnich and M. Henneaux
{\em Isomorphisms between the Batalin-Vilkovisky antibracket and the
  Poisson bracket}, J.Math.Phys. {\bf 37} (1996) 5273--5296.

\bibitem[GeDi1]{GeDi1}
I.M. Gel'fand and L.A. Dickey, {\em Lie algebra structure in the
  formal variational calculus}, Funkz. Anal. Priloz. {\bf 10 no~1}
  (1976), 18--25.

\bibitem[GeDi2]{GeDi2}
I.M. Gel'fand and L.A. Dickey, {\em Fractional powers of operators and
  hamiltonian systems}, Funkz. Anal. Priloz. {\bf 10 no~4} (1976), 13--29.

\bibitem[GeDo1]{GeDo1}
I.M. Gel'fand and I.~Ya. Dorfman, {\em Hamiltonian operators and associated
  algebraic structures}, Funkz. Anal. Priloz. {\bf 13 no~3} (1979), 13--30.

\bibitem[GeDo2]{GeDo2}
I.M. Gel'fand and I.~Ya. Dorfman, {\em Schouten bracket and {H}amiltonian operators}, Funkz. Anal.
  Priloz. {\bf 14 no~3} (1980), 71--74.

\bibitem[GeDo3]{GeDo3}
I.M. Gel'fand and I.~Ya. Dorfman, {\em Hamiltonian operators and infinite dimensional {L}ie 
algebras},
  Funkz. Anal. Priloz. {\bf 15 no~3} (1981), 23--40.

\bibitem[Dic1]{Dic1}
L.A. Dickey, {\em Soliton {E}quations and {H}amiltonian {S}ystems}, 
Advanced
  Series in Mathematical Physics, vol.~12, World Scientific, 1991.

\bibitem[Olv]{Olv}
P.J. Olver, {\em Applications of {L}ie {G}roups to {D}ifferential 
{E}quations},
  Graduate Texts in {M}athematics, vol. 107, Springer-Verlag, 1986.

\bibitem[Dic2]{Dic2}
L.A. Dickey, {\em Poisson brackets with divergence terms in field
  theories: two
  examples}, preprint, University of Oklahoma, 1997.


\bibitem[LaSt]{LaSt}
T.~Lada and J.D. Stasheff, {\em Introduction to sh {L}ie algebras for
  physicists}, Intern'l J. Theor. Phys. {\bf 32} (1993), 1087--1103.

\bibitem[LaMa]{LaMa}
T.~Lada and M.~Markl, {\em Strongly homotopy Lie algebras}, 
Comm.~in Algebra
  (1995), 2147--2161.


\bibitem[Kje]{Kje}
L.~Kjeseth, {\em {BRST} cohomology and homotopy {L}ie-{R}inehart pairs},
  Dissertation, UNC-CH, 1996.

\bibitem[Mar]{Mar} M. Markl, private communication.


\bibitem[LiZu]{LiZu} B. H. Lian andG. J. Zuckerman, 
{\em Commun. Math. Phys.} {\bf 154} (1993) 613--646.

\bibitem[GeVo]{GeVo} M. Gerstenhaber and A. A.Voronov, 
{\em Internat. Math. Research Notices} (1995) 141-153

\bibitem[KVZ]{KVZ} T. Kimura, A. A. Voronov, G. J. Zuckerman
     to appear in "Operads: Proceedings of Renaissance Conferences",
     J.-L. Loday, J. Stasheff, and A.A. Voronov, eds., {\em
     Contemp. Math.}, AMS, Providence, RI
     (1996).

\bibitem[BBH]{BBH} G. Barnich, F. Brandt and M. Henneuax, {\em Local BRST
    cohomology in the antifield formalism. 1. General theorems.},
Commun.Math.Phys. {\bf 174} (1995) 57--92.


\end{thebibliography}

\end{document}